\documentclass[prl,twocolumn,superscriptaddress]{revtex4}
\usepackage{graphics}           % for eps figures
\usepackage{bm}                 % for bold math symbols
\usepackage{amsmath}            % for \text and such
\usepackage{amssymb}
\usepackage{verbatim}           % useful for commenting out stuff
\usepackage{amsthm}             % does theorems etc. nicely, but
\theoremstyle{plain}            % This is the default

\def\bra#1{{\langle#1|}}
\def\ket#1{{|#1\rangle}}

\begin{document}

\title{Fermionic and bosonic quantum field theories from quantum cellular automata in three spatial dimensions}

\author{Leonard \surname{Mlodinow}}\email{lmlodinow@gmail.com}
\author{Todd A. \surname{Brun}}\email{tbrun@usc.edu}
\affiliation{Center for Quantum Information Science and Technology, University of Southern California, Los Angeles, California}

\date{\today}

%%% abstract before title in revtex4 %%%
\begin{abstract}
Quantum walks on lattices can give rise to relativistic wave equations in the long-wavelength limit, but going beyond the single-particle case has proven challenging, especially in more than one spatial dimension. We construct quantum cellular automata for distinguishable particles based on two different quantum walks, and show that by restricting to the antisymmetric and symmetric subspaces, respectively, a multiparticle theory for free fermions and bosons in three spatial dimensions can be produced. This construction evades a no-go theorem that prohibits the usual fermionization constructions in more than one spatial dimension. In the long-wavelength limit, these recover Dirac field theory and Maxwell field theory, i.e., free QED.
\end{abstract}


\maketitle

%\section{Introduction}

{\it Introduction}. Quantum field theories are traditionally constructed by starting with a classical field and then quantizing it. In the last decade or so, there have been a number of papers suggesting a new approach, based on a postulated QCA/QFT correspondence. That correspondence relates quantum cellular automata (QCAs) to quantum field theories (QFTs) \cite{Bialynicki94,Watrous95,Meyer96,Bracken07,Chandrashekar10,Yepez13,DAriano14,Arrighi14,Farrelly14a,Farrelly14b,Bisio15,Yepez16,Raynal17,MlodinowBrun18,BrunMlodinow19,Farrelly19,Arrighi20,MlodinowBrun20,BrunMlodinow20}.

The qubit is the fundamental unit of quantum information and quantum information processing is essentially the action of a string of unitary quantum gates on some initial state of qubits. In quantum field theory the time development of a quantum field is given by the action of a unitary operator on a state describing quantum particles, or the creation and annihilation operators corresponding to them. The QCA/QFT correspondence suggests that systems of the former type in the continuum limit yield familiar systems of the latter type. The idea is that one can construct a QCA from a quantum random walk based on a simple set of principles and symmetries, and then recover the desired Lorentz invariant QFT in the limit of continuous time and energies that are low enough that the lattice spacing is not probed (say the Planck length).

Previous work has largely been confined to fermion QFTs, and to the one-particle sector or the multiparticle sector, but in a single space dimension. Recent work has shown how fermion field theories can also be derived in two space dimensions \cite{BrunMlodinow20}. Here we present a derivation of both fermion and massless boson QFTs in 3+1 dimensions---a cellular automaton derivation of free QED. This correspondence to a QCA on a lattice with finite spacing provides a novel approach to avoiding the mathematical problems, such as infinities, of quantum field theory---one can view the emerging QFT as an effective theory of the underlying QCA. It also elucidates concretely the idea that information is physical, showing how one can make precise the somewhat vague but insightful idea of Wheeler \cite{Wheeler90}, ``it from bit''---the QCA/QFT correspondence implies, in fact, that the universe is a kind of quantum computer.

{\it Quantum walks}. A quantum walk is defined on a space $\mathcal{H}_{\rm QW} = \mathcal{H}_{\rm pos} \otimes \mathcal{H}_{\rm coin}$, where $\mathcal{H}_{\rm pos}$ is the space of possible positions of the particle, and $\mathcal{H}_{\rm coin}$ is an internal degree of freedom of the particle, commonly called the ``coin'' space, which indicates the direction in which the particle will move. (The term ``coin'' comes from the analogy with classical random walks, where a coin flip determines the direction that the particle moves.) The position space is spanned by a set of basis states $\{\ket{\mathbf{x}}\}$, where the positions $\mathbf{x}$ are the vertices of a graph. For this paper, we will assume that these vertices form a cubic lattice in three dimensions with lattice spacing $\Delta x$. The positions are $\mathbf{x} = (x,y,z) = (i\Delta x, j\Delta x, k\Delta x)$, with $i$, $j$ and $k$ integers. (For the present we will assume that this lattice has finite size $N$ with periodic boundary conditions.) The coin space has dimension $D$ (which will be different in different cases).

The general form we will use for the evolution of a quantum walk is
\begin{eqnarray}
U_{\rm QW} &=& e^{i\theta Q} \left( S_X P^+_X + S_X^\dagger P^-_X \right)
\left( S_Y P^+_Y + S_Y^\dagger P^-_Y \right) \nonumber\\
&& \times \left( S_Z P^+_Z + S_Z^\dagger P^-_Z \right) ,
\label{eq:QWalk3DForm}
\end{eqnarray}
where these $P^\pm$ operators form pairs of orthogonal projectors acting on the coin space: $P_X^+ + P_X^- = P_Y^+ + P_Y^- = P_Z^+ + P_Z^- = I$. $Q$ is an Hermitian operator on the coin space, called the ``coin flip operator,'' which obeys $Q^2=I$, and $\theta$ is a real parameter (which in the long-wavelength limit we will take to be small). $S_{X,Y,Z}$ are operators that act on the position space, and shift the particle by $+\Delta x$ in the $X$, $Y$, or $Z$ direction, respectively.   (The ordering of $X$, $Y$ and $Z$ in this expression is obviously arbitrary.)

Refs.~\cite{Chandrashekar11,Chandrashekar13} show that a quantum walk of the form of Eq.~(\ref{eq:QWalk3DForm}) can give nontrivial dynamics even if the coin flip operator is absent (corresponding to $\theta=0$ in our description here), so long as the projectors on the internal space corresponding to steps in different directions are noncommuting.  In \cite{MlodinowBrun18}, we argued that if a quantum walk is {\it unbiased} and {\it symmetric} then these operators must satisfy the following noncommutation relations:
\begin{eqnarray}
\{\Delta P_X, \Delta P_Y\} &=& \{\Delta P_X, \Delta P_Z\} = \{\Delta P_Y,\Delta P_Z\} \nonumber\\
&=& \{\Delta P_{X,Y,Z} , Q \} = 0 ,
\label{eq:deltaPs}
\end{eqnarray}
where $\Delta P_{X,Y,Z} = P_{X,Y,Z}^+ - P_{X,Y,Z}^-$. Four mutually anticommuting operators require a coin space of dimension $D\ge4$; we use $4\times4$ matrices $Q = \gamma_0$, $\Delta P_X = \gamma_0\gamma_1$, $\Delta P_Y = \gamma_0\gamma_2$, $\Delta P_Z = \gamma_0\gamma_3$, where the $\{\gamma_i\}$ are the gamma matrices.

{\it Momentum and energy representations}. We can change to a momentum representation,
\begin{equation}
\ket{\mathbf{k}} = \frac{1}{N^{3/2}} \sum_{\mathbf{x}} e^{-i\mathbf{k}\cdot\mathbf{x}} \ket{\mathbf{x}} ,
\end{equation}
where $\mathbf{k} =  (k_X, k_Y, k_Z) = (n\Delta k, o\Delta k, p\Delta k)$, where $\Delta k = 2\pi/(N\Delta x)$ and $n$, $o$, $p$ are integers. (We will take the range of these integers to be $-N/2+1$ to $N/2$ so that it makes sense to talk about momentum values close to zero.) The state $\ket{\mathbf{k}}$ is an eigenstate of the shifts, $S_{X,Y,Z}\ket{\mathbf{k}} = e^{i k_{X,Y,Z} \Delta x} \ket{\mathbf{k}}$. In the momentum representation the evolution is
\begin{eqnarray}
\label{eq:UQWmomentum}
U_{\rm QW} &=& e^{i\theta Q} e^{i \left(K_Z \Delta P_Z\right)\Delta x} \\
&& \times e^{i \left(K_Y \Delta P_Y\right)\Delta x} e^{i \left(K_X \Delta P_X\right)\Delta x}, \nonumber\\
&=& \sum_{\mathbf{k}} \ket{\mathbf{k}}\bra{\mathbf{k}} \otimes e^{i\theta Q} e^{i \left(k_Z \Delta P_Z\right)\Delta x} \nonumber\\
&& \times e^{i \left(k_Y \Delta P_Y\right)\Delta x} e^{i \left(k_X \Delta P_X\right)\Delta x} \nonumber\\
&\equiv& \sum_{\mathbf{k}} \ket{\mathbf{k}}\bra{\mathbf{k}} \otimes U_{\mathbf{k}}
\equiv  \sum_{\mathbf{k}} \ket{\mathbf{k}}\bra{\mathbf{k}} \otimes V_{\mathbf{k}} \Lambda_{\mathbf{k}} V_{\mathbf{k}}^\dagger , \nonumber
\end{eqnarray}
where $K_{X,Y,Z}$ are the operators corresponding to the components of the momentum and $\Lambda_{\mathbf{k}}$ are the diagonalized forms of $U_{\mathbf{k}}$.
%\begin{eqnarray}
%K_X \ket{\mathbf{k}} &=& k_X \ket{\mathbf{k}} , \nonumber\\
%K_Y \ket{\mathbf{k}} &=& k_Y \ket{\mathbf{k}} , \nonumber\\
%K_Z \ket{\mathbf{k}} &=& k_Z \ket{\mathbf{k}} .
%\end{eqnarray}
In this representation we see that the Hilbert space decomposes into $D$-dimensional blocks labeled by the momentum $\mathbf{k}$. The energy eigenstates $\ket{\mathbf{k}}\otimes\ket{\lambda_{{\mathbf k},j}}$ of the quantum walk can be found by diagonalizing the $D\times D$ matrices $U_{\mathbf{k}} = V_{\mathbf{k}} \Lambda_{\mathbf{k}} V_{\mathbf{k}}^\dagger$. We denote the eigenvalues by $\lambda_{{\mathbf k},j} \equiv e^{i\phi_{{\mathbf k}j}}$.

{\it QCAs for distinguishable particles}. As shown in \cite{BrunMlodinow20}, it is possible to embed a theory with up to $N_{\rm max}$ distinguishable particles into a QCA. This construction evades a no-go theorem \cite{MlodinowBrun20} that prohibits the usual fermionization constructions in more than one spatial dimension. The Hilbert space for the distinguishable particles is $\mathcal{H}_{\rm total} = \mathcal{H}^{(1)} \otimes \cdots \otimes \mathcal{H}^{(N_{\rm max})}$. $\mathcal{H}^{(j)} = \mathcal{H}_{\rm QW} \oplus {\rm span}(\ket\omega)$ is the Hilbert space for particle type $j$, which contains either one quantum walk particle, or no particle (the vacuum state $\ket\omega$). The entire system evolves by the unitary
\begin{equation}
U = U^{(1)} \otimes \cdots \otimes U^{(N_{\rm max})} ,
\end{equation}
where $U^{(j)} = U_{\rm QW} + \ket\omega\bra\omega$. This evolution conserves particle number. A state with $n$ particles in energy eigenstates $\ket{\lambda_{\mathbf{k}_1, j_1}}, \ldots, \ket{\lambda_{\mathbf{k}_n, j_n}}$, and the remaining $N_{\rm max} - n$ particle types in the vacuum state $\ket\omega$, will be eigenvectors of $U$ with eigenvalue $e^{i \sum_{\ell = 1}^n \phi_{\mathbf{k}_\ell j_\ell}}$. This eigenvalue does not depend on the ``type'' labels of the $n$ particles; permuting the particle types does not change the form of the states or their eigenvalues. So without loss of generality, we will consider the case where all $n$-particle states contain particles of types $1,\ldots,n$, and where particle types $n+1,\ldots,N_{\rm max}$ are in the vacuum state $\ket\omega$.

{\it Antisymmetrization, creation and annihilation operators}. To extract a theory of fermions from this QCA we restrict ourselves to the totally antisymmetric subspace of $\mathcal{H}_{\rm total}$. We call this the ``physical subspace,'' which we denote
\begin{equation}
\mathcal{H}_{\rm phys} = \mathcal{H}_0 \oplus \mathcal{A}_1 \oplus \cdots \oplus \mathcal{A}_{N_{\rm max}} .
\end{equation}
The subspace $\mathcal{A}_n$ is the totally antisymmetric subspace with $n$ particles, and $\mathcal{H}_0$ is the one-dimensional vacuum space ${\rm span}(\ket\Omega)$, where $\ket\Omega = \ket\omega^{\otimes N_{\rm max}}$. We can define basis vectors for the subspaces $\mathcal{A}_n$ in terms of the energy eigenstates of the quantum walk:
\begin{eqnarray}
\ket\Omega &=& \ket\omega^{\otimes N_{\rm max}} , \nonumber\\
\ket{\mathbf{k},j} &=& (\ket{\mathbf{k}} \otimes \ket{\lambda_{\mathbf{k},j}}) \otimes \ket\omega^{\otimes N_{\rm max}-1} , \nonumber\\
\ket{\mathbf{k}_1,j_1; \mathbf{k}_2,j_2} &=& \frac{1}{\sqrt2} \bigl(\ket{\mathbf{k}_1} \otimes \ket{\lambda_{\mathbf{k}_1,j_1}} \otimes  \ket{\mathbf{k}_2} \otimes \ket{\lambda_{\mathbf{k}_2,j_2}} - \nonumber\\
&& \ket{\mathbf{k}_2} \otimes \ket{\lambda_{\mathbf{k}_2,j_2}} \otimes  \ket{\mathbf{k}_1} \otimes \ket{\lambda_{\mathbf{k}_1,j_1}} \bigr) \nonumber\\
&& \otimes \ket\omega^{\otimes N_{\rm max}-2} ,
\end{eqnarray}
and so forth, defining basis states $\ket{\mathbf{k}_1,j_1; \cdots; \mathbf{k}_n,j_n}$ for particle numbers $n$ up to $N_{\rm max}$. The particle states $\mathbf{k}_i,j_i$ must all be distinct (or the state will vanish under antisymmetrization); and to avoid ambiguity, the order of $\mathbf{k}_1,j_1; \cdots; \mathbf{k}_n,j_n$ should follow some established ordering convention (it is irrelevant what convention is used).

The state $\ket{\mathbf{k}_1,j_1; \cdots; \mathbf{k}_n,j_n}$ is an eigenstate of the evolution operator $U$ with eigenvalue $e^{i \sum_{\ell = 1}^n \phi_{\mathbf{k}_\ell j_\ell}}$. So we see that this antisymmetrized ``physical subspace'' is preserved by the time evolution; an antisymmetric initial state will evolve to be antisymmetric at all later times.

Given this basis, we can define a set of creation and annihilation operators that transform these energy basis vectors into each other,
\begin{equation}
a^\dagger_{\mathbf{k}_1,j_1} \cdots a^\dagger_{\mathbf{k}_1,j_1} \ket\Omega = \ket{\mathbf{k}_1,j_1; \cdots; \mathbf{k}_n,j_n} ,
\end{equation}
and that obey the usual anticommutation relations $\{a^\dagger_{\mathbf{k_1},j_1}, a^\dagger_{\mathbf{k_2},j_2}\} = \{a_{\mathbf{k_1},j_1}, a_{\mathbf{k_2},j_2}\} = 0$, $\{a^\dagger_{\mathbf{k_1},j_1}, a_{\mathbf{k_2},j_2}\} = \delta_{\mathbf{k}_1,\mathbf{k}_2} \delta_{j_1,j_2}$. Any annihilation operator $a_{\mathbf{k},j}$ annihilates the vacuum $\ket\Omega$, and any creation operator $a^\dagger_{\mathbf{k},j}$ annihilates any state with $N_{\rm max}$ particles. These creation and annihilation operators are clearly nonlocal, but in spite of this we can write the action of the local evolution $U$ in terms of these operators for any $\ket\Psi \in \mathcal{H}_{\rm phys}$:
\begin{equation}
U\ket\Psi = e^{i \sum_{\mathbf{k},j} \phi_{\mathbf{k},j} a^\dagger_{\mathbf{k},j} a_{\mathbf{k},j}} \ket\Psi .
\end{equation}
This gives us an effective time evolution for the operators:
\begin{equation}
U a_{\mathbf{k},j} U^\dagger = e^{-i\phi_{\mathbf{k},j}} a_{\mathbf{k},j} .
\label{eq:momentumTimeEvolution}
\end{equation}

Having defined creation and annihilation operators for the energy eigenstates, we can then define creation and annihilation operators $\tilde{a}^\dagger_{\mathbf{k},j}$ and $\tilde{a}_{\mathbf{k},j}$ in the momentum representation. We can write these as operator-valued column vectors $\mathbf{\tilde{a}_k}^\dagger$ and $\mathbf{\tilde{a}_k}$, where
\begin{equation}
\mathbf{\tilde{a}_k}^\dagger = \left(\begin{array}{c} \tilde{a}^\dagger_{\mathbf{k},1} \\ \vdots \\  \tilde{a}^\dagger_{\mathbf{k},d} \end{array}\right)
= V_{\mathbf{k}} \left(\begin{array}{c} {a}^\dagger_{\mathbf{k},1} \\ \vdots \\  {a}^\dagger_{\mathbf{k},d} \end{array}\right) ,
\end{equation}
using the matrices $V_{\mathbf{k}}$ from Eq.~(\ref{eq:UQWmomentum}). These vectors have an effective time evolution
\begin{equation}
U \mathbf{\tilde{a}_k}^\dagger U^\dagger = U_{\mathbf{k}} \mathbf{\tilde{a}_k}^\dagger , \ \ \ 
U \mathbf{\tilde{a}_k} U^\dagger = U_{\mathbf{k}}^* \mathbf{\tilde{a}_k} .
\end{equation}

{\it The long-wavelength limit}. The long-wavelength limit of this QCA is the regime when all the particle momenta are small, $|\mathbf{k}|\Delta x \ll 1$. In this limit we can expand the quantum walk evolution operator
\begin{eqnarray}
U_{\rm QW} &\approx& I + i\theta Q + i\Delta x K_X \Delta P_X \nonumber\\
&& +  i\Delta x K_Y \Delta P_Y +  i\Delta x K_Z \Delta P_Z \nonumber\\
&=& I + i\gamma_0( \theta I + i\Delta x K_X \gamma_1 \nonumber\\
&& +  i\Delta x K_Y \gamma_2 +  i\Delta x K_Z \gamma_3 ,
\label{eq:uQWlongwavelength}
\end{eqnarray}
using the choices for $\Delta P_{X,Y,Z}$ after Eq.~(\ref{eq:deltaPs}). Calculating the eigenvalues $\lambda_{\mathbf{k},j} = e^{i\phi_{\mathbf{k},j}}$ for the operator $U_{\rm QW}$ in Eq.~(\ref{eq:uQWlongwavelength}), we can relate the phases $\phi_{\mathbf{k},j}$ to an energy
\begin{equation}
E_{\mathbf{k},j} \equiv \hbar\phi_{\mathbf{k},j}/\Delta t \approx \pm \sqrt{c^2(p_X^2 + p_Y^2 + p_Z^2) + m^2 c^4} ,
\end{equation}
where $c = \Delta x/\Delta t$ plays the role of the speed of light, $\mathbf{p} = \hbar\mathbf{k}$ is the momentum, and $m = (\hbar\Delta t/\Delta x^2)\theta$ is the effective rest mass.

In this limit, and for $N_{\rm max}$ and the lattice size $N$ very large, we can expand Eq.~(\ref{eq:momentumTimeEvolution}) to get the usual evolution equation for Dirac field theory:
\begin{equation}
\partial_t \mathbf{\tilde{a}_k}^\dagger \equiv \frac{1}{\Delta t} \left(U \mathbf{\tilde{a}_k}^\dagger U^\dagger - \mathbf{\tilde{a}_k}^\dagger \right) 
\approx -\frac{ic}{\hbar} \gamma_0 \bigl( \mathbf{p}\cdot\text{\boldmath$\gamma$} \bigr) \mathbf{\tilde{a}_k}^\dagger .
\end{equation}

{\it Quantum walk for bosons}. We would like to make a similar construction for free bosons. By analogy with the fermionic case, we will use a quantum walk structure similar to Eq.~(\ref{eq:QWalk3DForm}), but with a different choice of internal space and projectors; and instead of constructing the QCA using the fully antisymmetric subspace, we will use the fully symmetric subspace. 

In the case of bosons, rather than having the particle always move in one direction or the other along each axis, we will allow the possibility that the particle can also remain at rest. The QW evolution operator becomes
\begin{eqnarray}
U_{\rm QW} &=& e^{i\theta Q} \left( S_X P^+_X  + P^0_X + S_X^\dagger P^-_X \right) \nonumber\\
&&\times \left( S_Y P^+_Y + P^0_Y+ S_Y^\dagger P^-_Y \right) \nonumber\\
&& \times \left( S_Z P^+_Z + P^0_Z + S_Z^\dagger P^-_Z \right) .
\label{eq:QWalk3DForm2}
\end{eqnarray}
Since there are now three distinct possible motions for each axis ($+$, $-$ and $0$), we need a coin space with at least three dimensions. For simplicity, we will consider the case where there is no coin operator $\theta=0$, which by analogy to the Dirac result we will call the {\it massless} case. It turns out that in that case, a three-dimensional internal space is sufficient.

For unitarity, we require that each set of projectors $P_j^k$ for $j=X,Y,Z$ and $k=-,0,+$ be orthogonal and add to the identity. We also need a generalization of the {\it noncorrelation} or {\it equal norm} condition from \cite{MlodinowBrun18}, so that moving a direction along one axis does not in general bias the way the walk moves along a different axis. For the projectors $P^\pm_{X,Y,Z}$ this condition is
\begin{equation}
P_j^k P_{j'}^{+} P_j^k = P_j^k P_{j'}^{-} P_j^k = c P_j^k ,
\end{equation}
where $j,j' = X,Y,Z$, $j'\ne j$, $k=\pm$, and $c>0$.

We want the amplitude for the particle to remain in its initial location without moving to be zero. This condition requires that $P_X^0 P_Y^0 P_Z^0 = 0$, which implies that either $P_X^0 P_Y^0 = 0$ or $P_Y^0 P_Z^0 = 0$ or both. Because we want to treat motion along all three axes similarly, and not have the form of the evolution operator be highly dependent on the order of the shifts, we choose the symmetric condition
\begin{equation}
P_X^0 P_Y^0 = P_Y^0 P_Z^0 = P_Z^0 P_X^0 = 0 .
\end{equation}
Finally, combining these projectors with the $P_j^\pm$ projectors we require
\begin{equation}
P_j^k P_{j'}^{0} P_j^k = c' P_j^k ,\ \ 
P_j^0 P_{j'}^\pm P_j^0 = c' P_j^0 ,
\end{equation}
where again $j,j' = X,Y,Z$, $j'\ne j$, $k=\pm$, and $c'$ is a positive real constant. This set of conditions is plausible, and (as we shall see) also leads to interesting dynamics in the long-wavelength limit.

Switching to the momentum picture the evolution is
\begin{eqnarray}
U_{\rm QW} &=& e^{i \left(K_Z \Delta P_Z\right)} e^{i \left(K_Y \Delta P_Y\right)\Delta x} e^{i \left(K_X \Delta P_X\right)\Delta x}, \\
&\equiv& \sum_{\mathbf{k}} \ket{\mathbf{k}}\bra{\mathbf{k}} \otimes U_{\mathbf{k}} 
\equiv  \sum_{\mathbf{k}} \ket{\mathbf{k}}\bra{\mathbf{k}} \otimes V_{\mathbf{k}} \Lambda_{\mathbf{k}} V_{\mathbf{k}}^\dagger , \nonumber
\label{eq:momentumQWBoson}
\end{eqnarray}
where $\Delta P_{X,Y,Z} = P_{X,Y,Z}^+ - P_{X,Y,Z}^-$. In this case, $(\Delta P_{X,Y,Z})^2 = I - P_{X,Y,Z}^0$. We can find a suitable set of operators using the spin-1 matrices: $\Delta P_{X,Y,Z} = J_{X,Y,Z}$,
\[
J_X = \left(\begin{array}{ccc} 0 & 0 & 0 \\ 0 & 0 & -i \\ 0 & i & 0 \end{array}\right) , \ \ \ 
J_Y = \left(\begin{array}{ccc} 0 & 0 & i \\ 0 & 0 & 0 \\ -i & 0 & 0 \end{array}\right) ,
\]
\begin{equation}
J_Z = \left(\begin{array}{ccc} 0 & -i & 0 \\ i & 0 & 0 \\ 0 & 0 & 0 \end{array}\right) ,
\end{equation}
with $P^\pm_{X,Y,Z} = (1/2)(J^2_{X,Y,Z} \pm J_{X,Y,Z})$ and $P_{X,Y,Z}^0 = I - J_{X,Y,Z}^2$. Just as we did in the fermion case, we can decompose the Hilbert space into $3\times3$ blocks with a given momentum vector $\mathbf{k}$ as shown in Eq.~(\ref{eq:momentumQWBoson}), and diagonalize the unitaries $U_{\mathbf{k}} = V_{\mathbf{k}} \Lambda_{\mathbf{k}} V_{\mathbf{k}}^\dagger$ that act on those blocks. The eigenvalues of $U_{\mathbf{k}}$ are the diagonal elements of the diagonalized matrix $\Lambda_{\mathbf{k}}$. We can again denote them as $\lambda_{\mathbf{k},j} = e^{i\phi_{\mathbf{k},j}}$.

{\it QCA with symmetrization.} Just as in the fermion case, we embed $N_{\rm max}$ distinguishable particle types into a QCA. We confine ourselves into a subspace containing no more than one particle of each type; but this time, instead of completely antisymmetrized states we use completely symmetrized states. We can again define a set of basis vectors for this space:
\begin{eqnarray}
\ket\Omega &=& \ket\omega^{\otimes N_{\rm max}} , \nonumber\\
\ket{\mathbf{k},j} &=& (\ket{\mathbf{k}} \otimes \ket{\lambda_{\mathbf{k},j}}) \otimes \ket\omega^{\otimes N_{\rm max}-1} , \nonumber\\
\ket{\mathbf{k}_1,j_1; \mathbf{k}_2,j_2} &=& \frac{1}{\sqrt2} \bigl(\ket{\mathbf{k}_1} \otimes \ket{\lambda_{\mathbf{k}_1,j_1}} \otimes  \ket{\mathbf{k}_2} \otimes \ket{\lambda_{\mathbf{k}_2,j_2}} + \nonumber\\
&& \ket{\mathbf{k}_2} \otimes \ket{\lambda_{\mathbf{k}_2,j_2}} \otimes  \ket{\mathbf{k}_1} \otimes \ket{\lambda_{\mathbf{k}_1,j_1}} \bigr) \nonumber\\
&& \otimes \ket\omega^{\otimes N_{\rm max}-2} ,
\end{eqnarray}
and so on, where to remove ambiguity we establish an ordering of the states $\mathbf{k},j$ and require $\mathbf{k}_1,j_1 \le \mathbf{k}_2,j_2 \le \cdots \le \mathbf{k}_N,j_N$. Unlike the fermion case, the same state can be repeated any number of times (up to $N_{\rm max}$).

As before, we can define creation and annihilation operators that transform between these basis vectors:
\begin{eqnarray}
&& \frac{1}{\sqrt{m_1!\cdots m_n!}} \left( a_{\mathbf{k}_1,j_1}^\dagger\right)^{m_1} \cdots \left( a_{\mathbf{k}_n,j_n}^\dagger\right)^{m_n} \ket\Omega \nonumber\\
&=& \ket{ \underbrace{\mathbf{k}_1,j_1; \cdots; \mathbf{k}_1,j_1}_{\text{$m_1$ times}}; \cdots; \underbrace{\mathbf{k}_n,j_n; \cdots; \mathbf{k}_n,j_n}_{\text{$m_n$ times}}} .
\end{eqnarray}
These will obey the usual commutation relations,
\begin{equation}
[a^\dagger_{\mathbf{k_1},j_1},a^\dagger_{\mathbf{k_2},j_2}] = [a_{\mathbf{k_1},j_1},a_{\mathbf{k_2},j_2}] = 0 ,
\end{equation}
\begin{equation}
[a^\dagger_{\mathbf{k_1},j_1},a_{\mathbf{k_2},j_2}] = \delta_{\mathbf{k}_1,\mathbf{k}_2} \delta_{j_1,j_2} .
\end{equation}
Their time evolution, as before, will be given by
\begin{equation}
U a_{\mathbf{k},j} U^\dagger = e^{-i\phi_{\mathbf{k},j}} a_{\mathbf{k},j} .
\end{equation}
As in the fermionic case, this can be used to define an evolution equation for bosonic operators $\mathbf{\tilde{a}_k}$ in the momentum representation.

%{\it The long-wavelength limit}. If we go to the long-wavvelength limit $|\mathbf{k}\Delta x| \ll 1$ then we can expand the matrix $U_{\mathbf{k}}$ to get
%\begin{equation}
%U_{\mathbf{k}} \approx I + i \left( k_X J_X + k_Y J_Y + k_Z J_Z \right)\Delta x .
%\end{equation}
%From this we can define a Hamiltonian operator $H = - c \mathbf{P}\cdot\mathbf{J}$, where $\mathbf{P} = \hbar(K_X,K_Y,K_Z)$, $\mathbf{J} = (J_X,J_Y,J_Z)$, and $c=\Delta x/\Delta t$. The Hamiltonian operator has eigenvalues $\pm \sqrt{c^2 (p_X^2 + p_Y^2 + p_Z^2)}$ and 0. The evolution equation for $\mathbf{\tilde{a}_k}$ becomes
%\begin{equation}
%\partial_t \mathbf{\tilde{a}_k} = i c (\mathbf{k}\cdot\mathbf{J}) \mathbf{\tilde{a}_k}
%= c \mathbf{k}\times\mathbf{\tilde{a}_k} ,
%\end{equation}
%which is equivalent to Maxwell quantum field theory in the momentum representation.

{\it The long-wavelength limit}. If we go to the long-wavelength limit $|\mathbf{k}\Delta x| \ll 1$ then we can expand the matrix $U_{\mathbf{k}}$ to get
\begin{equation}
U_{\mathbf{k}} \approx I + i \left( k_X J_X + k_Y J_Y + k_Z J_Z \right)\Delta x .
\end{equation}
From this we can define a Hamiltonian operator $H = - c \mathbf{P}\cdot\mathbf{J}$, where $\mathbf{P} = \hbar(K_X,K_Y,K_Z)$, $\mathbf{J} = (J_X,J_Y,J_Z)$, and $c=\Delta x/\Delta t$. The Hamiltonian operator has eigenvalues $\pm \sqrt{c^2 (p_X^2 + p_Y^2 + p_Z^2)}$ and 0 with complex eigenvectors $\mathbf{v}_\pm$ and $\mathbf{v}_0$. Restricting ourselves to the positive energy sector leaves only $\mathbf{v}_+$.

It is conventional, however to view the photon's internal space as a real space so we define $\mathbf{\hat{e}}_{\mathbf{k},1} = {\rm Re}(\mathbf{v}_+)$ and $\mathbf{\hat{e}}_{\mathbf{k},2} = {\rm Im}(\mathbf{v}_+)$, giving:
\begin{eqnarray}
\mathbf{\hat{e}}_{\mathbf{k},1} &\equiv& \frac{1}{k\sqrt{k^2 - k_X^2}} \left(\begin{array}{c} k^2 - k_X^2 \\
- k_X k_Y \\ - k_X k_Z \end{array}\right) ,\nonumber\\
\mathbf{\hat{e}}_{\mathbf{k},2} &\equiv& \frac{1}{\sqrt{k^2 - k_X^2}} \left(\begin{array}{c} 0 \\
k_Z \\ - k_Y \end{array}\right) ,
\end{eqnarray}
where $k=|\mathbf{k}| = \sqrt{k_X^2 + k_Y^2 + k_Z^2}$. These are the usual transverse polarization vectors. We define a new set of boson operators ${b}^\dagger_{\mathbf{k},j}$ and ${b}_{\mathbf{k},j}$ for $j=1,2$.  In terms of these operators the Hamiltonian is
\begin{equation}
H_{\rm photon} = \sum_{\mathbf{k}} \sum_{j=1}^2 c\hbar k {b}^\dagger_{\mathbf{k},j}
{b} _{\mathbf{k},j} .
\end{equation}
From these boson operators, one can construct the usual expressions for the vector potential $\mathbf{A}$ and the fields $\mathbf{E}$ and $\mathbf{B}$ (see, e.g., Ref.~\cite{DrummondHillery14}):
\begin{eqnarray}
\mathbf{A}(\mathbf{x},t) &=& \sum_{\mathbf{k},j} \sqrt{\frac{\hbar}{2\epsilon_0\omega_kV}}
  \mathbf{\hat{e}}_{\mathbf{k},j} \left( b_{\mathbf{k},j} e^{i\mathbf{k}\cdot\mathbf{x}} +
  b^\dagger_{\mathbf{k},j} e^{-i\mathbf{k}\cdot\mathbf{x}} \right) , \nonumber\\
\mathbf{E}(\mathbf{x},t) &=& \sum_{\mathbf{k},j} i \sqrt{\frac{\hbar\omega_k}{2\epsilon_0V}}
  \mathbf{\hat{e}}_{\mathbf{k},j} \left( b_{\mathbf{k},j} e^{i\mathbf{k}\cdot\mathbf{x}} -
  b^\dagger_{\mathbf{k},j} e^{-i\mathbf{k}\cdot\mathbf{x}} \right) ,
\end{eqnarray}
where $V$ is the total volume and $\mathbf{B} = \nabla\times\mathbf{A}$.

{\it Conclusions and future work.} We have presented a construction for quantum cellular automaton models, based on antisymmetrizing or symmetrizing over distinguishable particle types, which recovers Dirac's and Maxwell's quantum field theories in the long-wavelength limit. Many further challenges remain. We wish to go beyond free particles to include interactions. Interacting theories will most likely not conserve particle number, which may necessitate new techniques to preserve the local dynamics of a QCA without losing the fermionic (or bosonic) statistics of the theory.  The scalar boson and massive vector boson cases are also open questions. And it would be very interesting to calculate observable consequences of discrete spacetime in these multiparticle theories.

%\begin{figure}[t]
%\includegraphics{fig1.eps}
%\caption{\label{fig1}  The caption.}
%\end{figure}

%\section{Conclusions}

\begin{acknowledgments}

LM and TAB acknowledge the hospitality of Caltech's Institute for Quantum Information and Matter (IQIM) and USC's Center for Quantum Information Science and Technology (CQIST).

\end{acknowledgments}

%-------------------------------------%
%%%% construct refs with BibTeX... %%%%
%\bibliographystyle{amsplain}
%\bibliographystyle{apsrev}
%\bibliography{paper}

\begin{thebibliography}{99}

\bibitem{Bialynicki94} I. Bialynicki-Birula, {\sl Weyl, Dirac, and Maxwell equations on a lattice as unitary cellular automata}, Phys. Rev. D {\bf 49}, 6920 (1994).

\bibitem{Watrous95} J. Watrous, {\sl On one-dimensional quantum cellular automata}, in {\sl Proceedings of the 36th Annual Symposium on Foundations of Computer Science}, 528--537 (1995). 

\bibitem{Meyer96} D.A. Meyer, {\sl From quantum cellular automata to quantum lattice gases}, J. Stat. Phys. {\bf 85}, 551--574 (1996).

\bibitem{Bracken07} A. J. Bracken, D. Ellinas and I. Smyrnakis, {\sl Free-Dirac-particle evolution as a quantum random walk}, Phys. Rev. A {\bf 75}, 022322 (2007).

\bibitem{Chandrashekar10} C. M. Chandrashekar, S. Banerjee and R. Srikanth, {\sl Relationship between quantum walks and relativistic quantum mechanics}, Phys. Rev. A {\bf 81}, 062340 (2010).

\bibitem{Yepez13} J. Yepez, {\sl Quantum Lattice Gas Model of Dirac Particles in 1+1 Dimensions}, arXiv:1307.3595.

\bibitem{DAriano14} G. M. D'Ariano and P. Perinotti, {\sl Derivation of the Dirac equation from principles of information processing}, Phys. Rev. A {\bf 90}, 062106 (2014).

\bibitem{Arrighi14} P. Arrighi, M. Forets and Vincent Nesme, {\sl The Dirac equation as a quantum walk: higher dimensions, observational convergence}, J. Phys. A {\bf 47}, 465302 (2014).

\bibitem{Farrelly14a} T. C. Farrelly and A. J. Short, {\sl Causal fermions in discrete space-time}, Phys. Rev. A {\bf 89}, 012302 (2014).

\bibitem{Farrelly14b} T. C. Farrelly and A. J. Short, {\sl Discrete spacetime and relativistic quantum particles}, Phys. Rev. A {\bf 89}, 062109 (2014).

\bibitem{Bisio15} A. Bisio, G.M. D’Ariano, P. Perinotti and A. Tosini, {\sl Weyl, Dirac and Maxwell quantum cellular automata}, Found. Phys. {\bf 45}, 1203--1221 (2015).

\bibitem{Yepez16} J. Yepez, {\sl Quantum Lattice Gas Algorithmic Representation of Gauge Field Theory}, Proc. SPIE 9996, {\sl Quantum Information Science and Technology II}, 99960N (24 October 2016).

\bibitem{Raynal17} Phillippe Raynal, {\sl Simple derivation of the Weyl and Dirac quantum cellular automata}, Phys. Rev. A {\bf  95}, 062344 (2017).

\bibitem{MlodinowBrun18} L. Mlodinow and T.A. Brun, {\sl Discrete spacetime, quantum walks and relativistic wave equations}, Phys. Rev. A {\bf 97}, 042131 (2018).

\bibitem{BrunMlodinow19} T.A. Brun and L. Mlodinow, {\sl Detection of discrete spacetime by matter interferometry}, Phys. Rev. D {\bf 99}, 015012 (2019).

\bibitem{Farrelly19} T. Farrelly, {\sl A Review of Quantum Cellular Automata}, arXiv:1904.13318.

\bibitem{Arrighi20} P. Arrighi, C. B\'eny and T. Farrelly, {\sl A quantum cellular automaton for one-dimensional QED}, Quant. Inf. Proc. {\bf 19}, 88 (2020).

\bibitem{MlodinowBrun20} L. Mlodinow and T.A. Brun, {\sl Quantum field theory from a quantum cellular automaton in one spatial dimension and a no-go theorem in higher dimensions}, to appear in Phys. Rev. A, arXiv:2006.08927.

\bibitem{BrunMlodinow20} T.A. Brun and L. Mlodinow, {\sl Quantum cellular automata and quantum field theory in two spatial dimensions}, submitted to Phys. Rev. A.

\bibitem{Wheeler90} J.A. Wheeler, {\sl Information, Physics, Quantum: The Search For Links}, in {\sl Symp. Foundations of Quantum Mechanics, Tokyo 1989}, pp 354--368.

\bibitem{Chandrashekar11} C.M. Chandrashekar, {\sl Two-state quantum walk on two- and three-dimensional lattices}, arXiv:1103.2704.

\bibitem{Chandrashekar13} C.M. Chandrashekar, {\sl Two-component Dirac-like Hamiltonian for generating quantum walk on one-, two- and three-dimensional lattices}, Scientific Reports {\bf 3}, 2829 (2013).

\bibitem{DrummondHillery14} P.D. Drummond and M. Hillery. {\sl The quantum theory of nonlinear optics} (Cambridge University Press, Cambridge, 2014).

\end{thebibliography}
%-------------------------------------%

\end{document}